\documentclass[useAMS,usenatbib,psfig]{mn2e}
\usepackage{psfig}
\usepackage{times}

\title[An accretion disc-corona model for X-ray spectra of AGN]
{An accretion disc-corona model for X-ray spectra of active galactic
nuclei}
\author[X. Cao]
{ Xinwu Cao\\
Shanghai Astronomical Observatory, Chinese Academy of Sciences, 80
Nandan Road, Shanghai, 200030, China; E-mail: cxw@shao.ac.cn }

\date{Accepted 2008 December 9.  Received 2008 December 8; in original form 2008 September 15}

\pagerange{\pageref{firstpage}--\pageref{lastpage}} \pubyear{}

\begin{document}

\maketitle \label{firstpage}

\begin{abstract}
The hard X-ray emission of active galactic nuclei (AGN) is believed
to originate from the hot coronae above the cold accretion discs.
The hard X-ray spectral index is found to be correlated with the
Eddington ratio $L_{\rm bol}/L_{\rm Edd}$, and the hard X-ray
bolometric correction factor $L_{\rm bol}/L_{\rm X,2-10keV}$
increases with the Eddington ratio. {The Compton reflection is also
found to be correlated with the hard X-ray spectral index for
Seyfert galaxies and X-ray binaries.} These observational features
provide very useful constraints on the accretion disc-corona model
for AGN. We construct an accretion disc-corona model and calculate
the spectra with different magnetic stress tensors in the cold
discs, in which the corona is assumed to be heated by the
reconnection of the magnetic fields generated by buoyancy
instability in the cold accretion disc. Our calculations show that
the magnetic stress tensor $\tau_{r\varphi}=\alpha p_{\rm gas}$
fails to explain all these observational features, while the
disc-corona model with $\tau_{r\varphi}=\alpha p_{\rm tot}$ always
leads to constant $L_{\rm bol}/L_{\rm X,2-10keV}$ independent of the
Eddington ratio. The resulted spectra of the disc-corona systems
with $\tau_{r\varphi}=\alpha \sqrt{p_{\rm gas}p_{\rm tot}}$ show
that both the hard X-ray spectral index and the hard X-ray
bolometric correction factor $L_{\rm bol}/L_{\rm X,2-10keV}$
increase with the Eddington ratio, which are qualitatively
consistent with the observations. We find that the disc-corona model
is unable to reproduce the observed very hard X-ray continuum
emission from the sources accreting at low rates (e.g., $\Gamma\sim
1$ for $L_{\rm bol}/L_{\rm Edd}\sim 0.01$), which may imply the
different accretion mode in these low luminosity sources. We suggest
that the disc-corona system transits to an advection-dominated
accretion flow$+$disc corona system at low accretion rates, which
may be able to explain all the above-mentioned correlations.
\end{abstract}


\section{Introduction}

It is believed that active galactic nuclei (AGN) are powered by
accretion of matter on to massive black holes, and the observed
UV/optical emission of AGN is thought to be a thermal emission from
the standard geometrically thin, optically thick accretion discs
\citep*[e.g.,][]{s78,ms82,sm89}. On the other hand, the power-law
hard X-ray spectra of AGN are most likely due to the inverse-Compton
scattering of soft photons on a population of hot electrons
\citep{grv79,hm91,hm93}. In the accretion disc-corona model, such
soft photons are from the cold disc, a fraction of which are Compton
scattered by the hot electrons in the corona above the cold disc to
the hard X-ray energy band. The disc-corona model was extensively
explored in many previous works
\citep*[e.g.,][]{sz94,hm91,hm93,ksm01,lms02,lmo03}. In this
disc-corona scenario, most gravitational energy is generated in the
cold disc, probably through turbulence produced by the
magneto-rotational instability \citep{bh91}. The magnetic fields
generated in the cold disc are strongly buoyant, and a substantial
fraction of magnetic energy is transported vertically to heat the
corona above the disc with the reconnection of the fields
\citep*[e.g.,][]{d98,dcf99,mf01,mf02}. It was found that the
temperature of the hot electrons in the corona is roughly around
$10^9$~K, which can successfully reproduce a power-law hard X-ray
spectrum as observed \citep*[e.g.,][]{lmo03}.

The physical processes of turbulence triggered by the amplified
magnetic fields in the disc are very complicated, and quite unclear,
though they are revealed to some extent with numerical
magneto-hydrodynamical (MHD) simulations \citep{bh91,bh98}. The
so-called ``$\alpha$-prescription" is widely adopted in most of the
works on accretion discs \citep{ss73}, in which the magnetic stress
tensor is assumed to be proportional to the total pressure ($p_{\rm
tot}=p_{\rm gas}+p_{\rm rad}$), gas pressure $p_{\rm gas}$, or
$\sqrt{p_{\rm gas}p_{\rm tot}}$ \citep{sc81,sr84,tl84}. The magnetic
stress tensor $\tau_{r\varphi}=\alpha p_{\rm tot}$ was suggested in
\citet{ss73}, however, the disc is thermal unstable if the radiation
pressure dominates over the gas pressure \citep{ss76}\citep*[but
also see][]{hkb08}. It was argued that the magnetic field
amplification is likely to be limited to that the magnetic pressure
is less than the gas pressure even in the inner
radiation-pressure-dominated regions, i.e.,
$\tau_{r\varphi}=\alpha{p}_{\rm gas}$ \citep{sc81}. The disc with
this stress tensor is thermal stable \citep{sc81}. The stress tensor
$\tau_{r\varphi}=\alpha \sqrt{p_{\rm gas}p_{\rm tot}}$ was initially
suggested by \citet{tl84} based on the idea that the viscosity is
proportional to the gas pressure, but the size of turbulence should
be limited by the disc scaleheight that is given in terms of the
total pressure. This is also supported by the analysis on the local
dynamical instabilities in magnetized, radiation-pressure-supported
accretion discs \citep{bs01}, which also leads to $p_{\rm
m}=B^2/8\pi\simeq \beta_0 \sqrt{p_{\rm gas}p_{\rm tot}}$, where
$p_{\rm m}$ is the magnetic pressure, and $\beta_0$ is a constant of
the order of unity \citep*[see][for a detailed discussion]{mf02}.
\citet{nrm00} argued that the accretion disc model with
$\tau_{r\varphi}=\alpha p_{\rm gas}$ is too stable to explain the
unstable behaviour observed in GRS~1915+105, and they further
suggested that it is possible to couple the radiation to the
particles through collisions and thereby allow the radiation
pressure to contribute to the stress tensor to some extent. As the
complexity of the physics in the radiation-pressure-dominated
fluids, any one of these stress tensors should only be regarded as a
possible option in accretion discs.

The hard X-ray observations on AGN may provide useful clues on the
accretion disc-corona models. \citet{wwm04} compiled a sample of
radio-quite AGN, and found a strong correlation between $L_{\rm
2-10keV}/L_{\rm bol}$ and $L_{\rm bol}/L_{\rm Edd}$, which was
conformed by \citet{vf07} with a different AGN sample. This
correlation was then used to constrain the disc-corona models with
different magnetic stress tensors, and they found that the model
with magnetic stress tenor $\tau_{r \varphi}=\alpha p_{\rm gas}$ is
favored by the correlation of $L_{\rm 2-10keV}/L_{\rm bol}-L_{\rm
bol}/L_{\rm Edd}$. It was also found that the spectral index of hard
X-ray continuum emission is correlated with the Eddington ratio
\citep{ly99,wwm04,s06,s08}. {\citet{zls99} found a strong
correlation between the Compton reflection and the hard X-ray
spectral index for Seyfert galaxies and X-ray binaries.} In the
accretion disc-corona scenario, the hard X-ray emission originates
from the Compton scattering of the soft photons by the hot electrons
in the corona. Therefore, the correlations of the hard X-ray
spectral index with the Eddington ratio/Compton reflection, together
with the correlation of $L_{\rm 2-10keV}/L_{\rm bol}-L_{\rm
bol}/L_{\rm Edd}$, provide important constraints on the accretion
disc-corona model.

{The parallel plane homogeneous corona is unable to produce an X-ray
spectrum with $\Gamma<2$ in 2--10~keV. The X-ray photons radiated
from the corona are reprocessed in the cold disc, and the
reprocessed photons irradiate the corona, which cool the corona and
lead to rather soft X-ray spectra with $\Gamma\ga 2$. In the patchy
corona model proposed by \citet{hmg94}, the corona appears as
individual blobs above the cold disc. Most of the reprocessed
photons do not enter the blob, so the cooling is significantly
reduced and the blobs are hotter than the parallel plane homogeneous
corona, which leads to harder X-ray spectra. This model can explain
the observed hard X-ray spectra with $\Gamma<2$ in some AGN
\citep*[e.g.,][]{z96}, however, it is still unable to explain the
correlation between the Compton reflection and the hard X-ray
spectral index for Seyfert galaxies and X-ray binaries
\citep{zls99}, which is also the case for the parallel plane
homogeneous corona model. \citet{zls99} suggested that a central hot
plasma surrounded by a cold disc may explain the correlation of the
Compton reflection with the hard X-ray spectral index. In this
scenario, the cold disk is truncated at a certain radius $d$, within
which the hot plasma may possibly correspond to an advection
dominated accretion flow (ADAF) \citep*[e.g.,][]{ny95}. The
radiation of the hot plasma is dominated by the inverse Compton
scattering of the soft photons from the outer cold disc, while the
cold disc is irradiated by the inner hot plasma. Thus, the Compton
reflection component decreases with increasing the inner radius of
the cold disc, which leads to less soft seed photons from the cold
disc entering the hot plasma and then the harder X-ray spectrum from
the hot plasma. The correlation between the Compton reflection and
the hard X-ray spectral index can be reproduced by this model
\citep*[see][for the details]{zls99}. An alternative model was
proposed by \citet{b99} to explain the correlation between the
Compton reflection and the X-ray spectral index. In this model, the
hot plasma above the cold disc is assumed to move away from the cold
disc at a mild relativistic velocity. Such an outflow reduces the
downward flux, and then reduces both the reflection and reprocessing
in the cold disc. The reduction of the reprocessing leads to less
incident soft seed photons and cooling in the hot plasma above the
cold disc, and in turn leads to harder X-ray spectrum
\citep{b99,mbp01}.  }

In this work, we take these different magnetic stress tensors as
candidates in our disc-corona model calculations, which could be
tested with the X-ray observations of AGN. We summarize the
disc-corona model in Sect. 2, and the numerical results of the model
calculations are given in Sect. 3. In Sect. 4, we discuss the
physical implications of the results.

\section{The disc-corona model}

The gravitational power dissipated in unit surface area of the
accretion disc is given by
\begin{equation}
Q_{\rm dissi}^+={\frac {3}{8\pi}}\dot{M}\Omega_{\rm
K}(R)^2\left[1-\left({\frac {R_{\rm in}}{R}}\right)^{1/2}\right],
\label{q_dissi}
\end{equation}
where $\dot{M}$ is the mass accretion rate of the disc, $\Omega_{\rm
K}(R)$ is the Keplerian velocity at radius $R$, and $R_{\rm
in}=3R_{\rm S}$ \citep{ss73}. The Schwarzschild radius $R_{\rm
S}=2GM_{\rm bh}/c^2$, where $M_{\rm bh}$ is the black hole mass. The
corona is assumed to be heated by the reconnection of the magnetic
fields generated by buoyancy instability in the disc.

The power dissipated in the corona is \citep{d98}
\begin{equation}
Q_{\rm cor}^{+}=p_{m}v_{\rm p}={\frac {B^2}{8\pi}}v_{\rm p},
\label{q_cor}
\end{equation}
where $p_{\rm m}$ is the magnetic pressure in the disc, and $v_{\rm
p}$ is the velocity of the magnetic flux transported vertically in
the disc. The rising speed $v_{\rm p}$ is assumed to be proportional
to their internal Alfven velocity, i.e., $v_{\rm p}=bv_{\rm A}$, in
which $b$ is of the order of unity for extremely evacuated magnetic
tubes.

The soft photons from the disc are Compton scattered by the hot
electrons in the corona to X-ray bands, and about half of the
scattered photons are intercepted by the disc. The reflection albedo
$a$ is relatively low, $a\sim0.1-0.2$, and most of the incident
photons from the corona are re-radiated as blackbody radiation
\citep*[e.g.,][]{zls99}. Thus, the energy equation for the cold disc
is
\begin{equation}
Q_{\rm dissi}^{+}-Q_{\rm cor}^{+}+{\frac {1}{2}}(1-a)Q_{\rm
cor}^{+}={\frac {4\sigma T_{\rm disc}^4}{3\tau}},
\label{disk_energy}
\end{equation}
where $T_{\rm disc}$ is the effective temperature in the mid-plane
of the disc, and $\tau=\tau_{\rm es}+\tau_{\rm ff}$ is the optical
depth in vertical direction of the disc. In this work, we adopt
$a=0.15$ in all our calculations.

As the detailed physics for generating magnetic fields in the
accretion disc is still quite unclear, we adopt different magnetic
stress tensors as:
\begin{equation}
 \tau_{r\varphi}=p_{\rm m}=\left\{ \begin{array}{ll}
        \alpha p_{\rm tot}; \\
         \alpha p_{\rm gas}; \\
         \alpha \sqrt{p_{\rm gas}p_{\rm tot}}, \\
\end{array} \right.
\label{viscosity}
        \end{equation}
in our model calculations, respectively. We summarize the equations
describing the disc as follows:

The continuity equation of the disc is
\begin{equation}
-4\pi RH_{\rm d}(R)\rho(R) v_{\rm R}(R)=\dot{M}, \label{continuity}
\end{equation}
where $H_{\rm d}(R)$ is the half thickness of the disc, $\rho(R)$ is
the mean density of the disc, and $v_{\rm R}(R)$ is the radial
velocity of the accretion flow at radius $R$.

The equation of state for the gas in the disc is
\begin{equation}
p_{\rm tot}=p_{\rm gas}+p_{\rm rad}={\frac {\rho kT_{\rm disc}}{\mu
m_{\rm p}}}+{\frac {1}{3}}aT_{\rm disc}^4, \label{state}
\end{equation}
{where $\mu=(1/\mu_{\rm i}+1/\mu_{\rm e})^{-1}$, $\mu_{\rm i}=1.23$
and $\mu_{\rm e}=1.14$ are adopted corresponding to the plasma
consisting of 3/4 hydrogen and 1/4 helium.} The vertical
hydrodynamical equilibrium requires $H_{\rm d}={c_{\rm
s}}/{\Omega_{\rm K}}$, where the sound speed $c_{\rm
s}=\sqrt{(p_{\rm tot}+p_{\rm m})/\rho}$.

The angular momentum equation for the disc is
\begin{equation}
\dot{M}\Omega_{\rm K}(R)\left[1-\left({\frac {R_{\rm
in}}{R}}\right)^{1/2}\right]=4\pi H_{\rm d}\tau_{r\varphi},
\label{angular}
\end{equation}
where the magnetic stress tensor is given by equation
(\ref{viscosity}).

Solving equations (\ref{q_dissi})-(\ref{angular}) numerically, the
structure of the disc and the power dissipated in the corona $Q_{\rm
cor}^{+}$ can be derived as functions of radius $R$. The ratio of
the power dissipated in the corona to the total for such a
disc-corona system is available:
\begin{equation}
\left<f\right>={\frac {\int Q_{\rm cor}^{+}2\pi R{\rm d}R}{\int
Q_{\rm dissi}^{+}2\pi R{\rm d}R}}. \label{f}
\end{equation}

The equation of state for the hot gas in the corona is
\begin{equation}
p_{\rm cor}={\frac {\rho_{\rm cor} kT_{\rm i}}{\mu_{\rm i}m_{\rm
p}}}+{\frac {\rho_{\rm cor} kT_{\rm e}}{\mu_{\rm e}m_{\rm
p}}}+p_{\rm cor,m}, \label{state_cor}
\end{equation}
where $T_{\rm i}$ and $T_{\rm e}$ are the temperatures of the ions
and electrons in the two-temperature corona, and {the magnetic
pressure $p_{\rm cor,m}=B_{\rm cor}^2/8\pi$. In this work, the
magnetic fields are assumed to be equipartition with the gas
pressure in the corona.} {The vertical hydrodynamical equilibrium in
the corona requires $H_{\rm cor}={c_{\rm cor,s}}/{\Omega_{\rm K}}$,
where the sound speed $c_{\rm cor,s}=\sqrt{p_{\rm cor}/\rho_{\rm
cor}}$. }

The energy equation describing the two-temperature corona is
\begin{equation}
Q_{\rm cor}^{+}=Q_{\rm cor}^{\rm ie}+\delta Q_{\rm cor}^{+}=F_{\rm
cor}^{-}, \label{cor_energy}
\end{equation}
where $F_{\rm cor}^{-}=F_{\rm syn}^{-}+F_{\rm brem}^{-}+F_{\rm
Comp}^{-}$ is the cooling rate in unit surface area of the corona
and $Q_{\rm cor}^{\rm ie}$ is the energy transfer rate from the ions
to the electrons in the corona via Coulomb collisions, which is
given by \citet{sg83}. The fraction of the energy directly heat the
electrons $\delta$ can be as high as $\sim 0.5$ by magnetic
reconnection, if the magnetic fields in the plasma are strong
\citep{bl97,bl00}. In this work, we adopt $\delta=0.5$ in our
calculations. {For the plasma consisting of different elements, the
Coulomb interaction between the electrons and ions is given by
\citep*[see][]{z98}
\begin{displaymath}
Q_{\rm cor}^{\rm ie}=1.5\sum_{Z}{\frac {m_{\rm e}}{m_{\rm
p}A_{Z}}}Z^{2}n_{\rm e}n_{\rm Z}H_{\rm cor}\sigma_{\rm T}c {\frac
{kT_{\rm i}-kT_{\rm e}}{K_2(1/\Theta_{\rm
e})K_2(1/\Theta_{Z})}}\ln\Lambda
\end{displaymath}
\begin{equation}
\times\left[{\frac {2(\Theta_{\rm e}+\Theta_{Z})^2+1}{\Theta_{\rm
e}+\Theta_{Z}}}K_{1}\left({\frac {\Theta_{\rm
e}+\Theta_{Z}}{\Theta_{\rm e}\Theta_{Z}}}\right)+2K_{0}\left({\frac
{\Theta_{\rm e}+\Theta_{Z}}{\Theta_{\rm
e}\Theta_{Z}}}\right)\right],\label{q_ie}
\end{equation}
where $A_Z$ is the mass number of the $Z$th element,
$\ln\Lambda=20$, $\Theta_{\rm e}=kT_{\rm e}/m_{\rm e}c^2$,
$\Theta_{Z}=\Theta_{\rm i}/A_Z$, and $\Theta_{\rm i}=kT_{\rm
i}/m_{\rm p}c^2$ are the dimensionless temperatures.} Although the
cooling rate of the corona is dominated by the Compton scattering of
the soft incident photons from the disc, we include the synchrotron,
bremsstrahlung and Compton emission in our calculations. The cooling
terms $F_{\rm syn}^{-}$ and $F_{\rm brem}^{-}$ are the functions of
electron number density, temperature, and the magnetic field
strength of the gas in the corona, which are taken from
\citet{ny95}.

{Assuming the corona to be a parallel plane, the mean probability of
the soft photons injected from the cold disc experiencing the
first-order scattering in the corona is
\begin{equation}
P_{1}=2\int\limits_{0}^{1}(1-e^{-\tau_0/\cos\theta})\cos\theta{\rm
d}\cos\theta,
 \label{prob_scat1}
\end{equation}
where the {constant specific intensity of the soft photons from the
disc is} assumed, $\theta$ is the angle of the motion of the soft
photons with respect to the vertical direction of the disc and
$\tau_0=\sigma_{T}n_{\rm e}H_{\rm cor}$ is the optical depth of the
corona for electron scattering in the vertical direction. {For
simplicity, we assume the first-order scattered photons are radiated
uniformly throughout the vertical direction of the corona.} The mean
probability for these {first-order} scattered photons experiencing
the second-order scattering can be estimated as
\begin{equation}
P_{2}={1\over2}\int\limits_{0}^{1}{\rm
d}\xi\int\limits_{0}^{1}[1-e^{-\xi\tau_0/\cos\theta}+1-e^{-(1-\xi)\tau_0/\cos\theta}]{\rm
d}\cos\theta, \label{prob_scat2}
\end{equation}
where $\xi=z/H_{\rm cor}$, and the {first-order} scattered photons
are assumed to be radiated isotropically. The probability of the
scattered photons experiencing next higher order scattering
approximates to $P_2$, so we simply adopt $P_{n}=P_2$ for $n>2$.}
Thus, using the method suggested by \citet{cb90}, we can calculate
the Comptonized spectrum with equations (\ref{prob_scat1}) and
(\ref{prob_scat2}), if the density, temperature of electrons, and
the incident spectrum of the disc are known. Integrating the derived
Comptonized spectrum over the frequency, the cooling of the
electrons in the corona due to the Compton scattering $F_{\rm
Comp}^{-}$ is available.

The disc structure and the power dissipated in the corona can be
calculated as functions of $R$, provided the black hole mass $M_{\rm
bh}$, mass accretion rate $\dot{M}$ and the viscosity parameter
$\alpha$ are specified. The scaleheight of the corona is a function
of $T_{\rm i}$, $T_{\rm e}$, and $n_{\rm e}$, on the assumption of
static hydrodynamical equilibrium in the vertical direction. There
are two equations describing the energy equilibrium in the corona at
$R$ (equation \ref{cor_energy}), which contain three physical
quantities of the corona: $T_{\rm i}$, $T_{\rm e}$, and $n_{\rm e}$.
\citet{lmo03} calculated the vertical structure of the
two-temperature corona above a thin disc based on the evaporation
mechanism, and their results show that the temperature of the ions
in the corona is always in the range of $\sim 0.2-0.3$ virial
temperature {(the virial temperature is defined as $T_{\rm
vir}=GMm_{\rm p}/kR$ in their work). In this work, we adopt a
slightly different definition: $T_{\rm vir}=GMm_{\rm p}/3kR$, as
that adopted by \citet{mf02}. To avoid the complexity of calculating
the vertical structure of the corona,} we simply adopt $T_{\rm
i}=0.9T_{\rm vir}$ [equivalent to $0.3$ in \citet{lmo03}'s work] in
our model calculations. Thus, the structure of the corona, i.e., the
temperature and density of the electrons, can be derived, and the
spectra of the disc-corona system are available based on the derived
disc-corona structure with different magnetic stress tensors given
in equation (\ref{viscosity}).

\section{Results}

\begin{figure}
\centerline{\psfig{figure=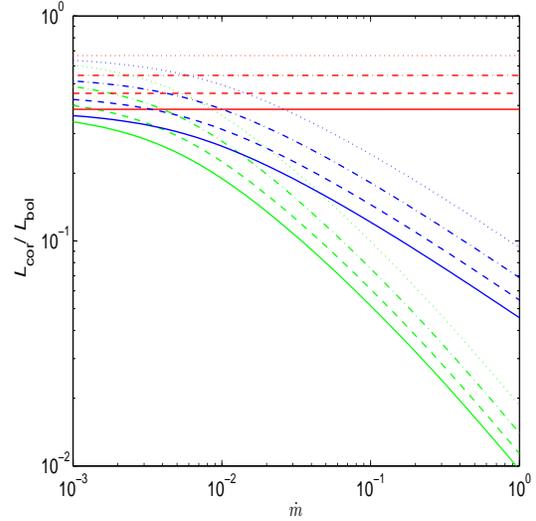,width=7.0cm,height=7.0 cm}}
\caption{The ratios $L_{\rm cor}/L_{\rm bol}$ as functions of
accretion rate $\dot{m}$ predicted by the models with different
magnetic stress tensors. The red lines represent the results
calculated with the magnetic stress tensor
$\tau_{r\varphi}=\alpha{p}_{\rm tot}$ (green lines:
$\tau_{r\varphi}=\alpha{p}_{\rm gas}$; blue lines:
$\tau_{r\varphi}=\alpha\sqrt{p_{\rm gas}p_{\rm tot}}$). The
different line types represent the different values of viscosity
parameter $\alpha$ adopted (solid lines: $\alpha=0.2$; dashed lines:
$\alpha=0.3$; dash-dotted lines: $\alpha=0.5$ and dotted lines:
$\alpha=1$). } \label{fig1}
\end{figure}

We calculate the disc-corona structure as described in section 2.
The black hole mass $M_{\rm bh}=10^8{\rm M}_\odot$ is adopted in all
our calculations, because the main features of the hard X-ray
spectra are almost independent of $M_{\rm bh}$ for massive black
holes in AGN. In Fig. \ref{fig1}, we plot the ratios of the power
radiated in the corona to the total power $L_{\rm cor}/L_{\rm bol}$
as functions of accretion rate $\dot{m}$ predicted by the models
with different magnetic stress tensors. {The dimensionless accretion
rate $\dot{m}$ is defined as $\dot{m}=\dot{M}/\dot{M}_{\rm Edd}$,
where $\dot{M}_{\rm Edd}=L_{\rm Edd}/\eta_{\rm eff}c^2$, and a
conventional radiative efficiency $\eta_{\rm eff}=0.1$ is adopted.}
The model with $\tau_{r\varphi}=\alpha{p}_{\rm tot}$ always leads to
constant ratios $L_{\rm cor}/L_{\rm bol}$ independent of accretion
rate $\dot{m}$, while the disc-corona model calculations with
$\tau_{r\varphi}=\alpha{p}_{\rm gas}$ show that $L_{\rm cor}/L_{\rm
bol}$ becomes extremely small for high accretion rates (e.g.,
$L_{\rm cor}/L_{\rm bol}\sim 0.01$ for $\dot{m}\sim 1$). The model
with $\tau_{r\varphi}=\alpha \sqrt{p_{\rm gas}p_{\rm tot}}$ shows
that the ratios $L_{\rm cor}/L_{\rm bol}\sim 0.3-0.6$ at
$\dot{m}=0.01$, while $L_{\rm cor}/L_{\rm bol}\sim 0.05-0.1$ at
$\dot{m}=1$, for different values of $\alpha$.

\begin{figure}
\centerline{\psfig{figure=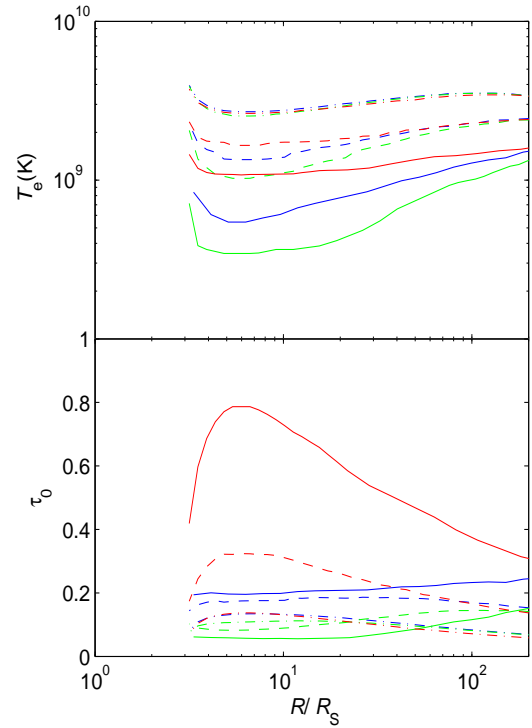,width=7.0cm,height=9.8 cm}}
\caption{The upper panel: the temperature of the electrons in the
corona as functions of disc radius $R$. The colours represent the
models with different magnetic stress tensors, which are the same as
those in Fig. \ref{fig1}. The solid lines represent the results
calculated for $\dot{m}=0.5$, while the dashed and dash-dotted lines
are for $\dot{m}=0.05$ and $0.005$, respectively. The viscosity
parameter $\alpha=0.5$ is adopted for all the model calculations. In
the lower panel, we plot the optical depth for the Compton
scattering of the electrons in the vertical direction of the corona.
} \label{fig2}
\end{figure}

The temperature and optical depth for Compton scattering of the hot
electrons in the vertical direction of the corona are plotted in
Fig. \ref{fig2}. The electron temperatures are in the range of
$\sim5\times 10^8-3\times 10^9$~K for different values of $\dot{m}$.
The temperature of the hot electrons in the corona tends to decrease
with accretion rate $\dot{m}$. When the accretion rate $\dot{m}$ is
as high as $\sim0.5$, the electron temperature of the corona
decreases to $\sim 5\times10^8$~K.

\begin{figure}
\centerline{\psfig{figure=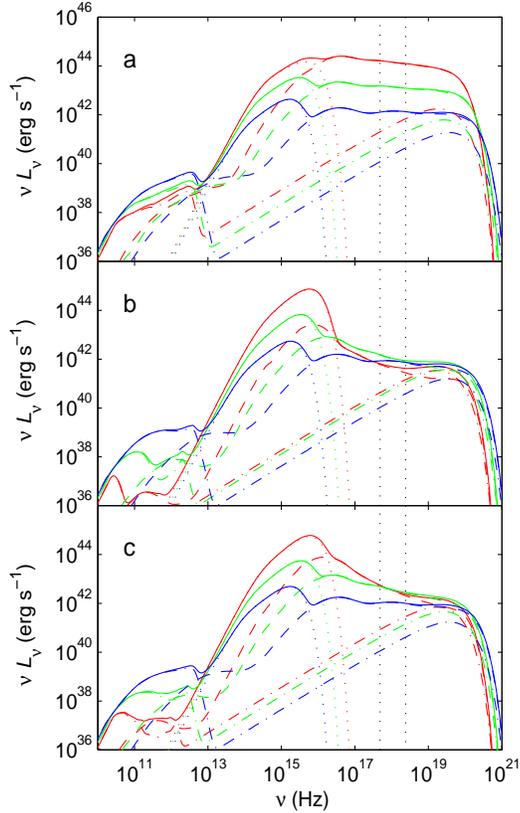,width=7.0cm,height=11.0 cm}}
\caption{The spectra of the disc-corona systems with different
magnetic stress tensors (a: $\tau_{r\varphi}=\alpha{p}_{\rm tot}$;
b: $\tau_{r\varphi}=\alpha{p}_{\rm gas}$; c:
$\tau_{r\varphi}=\alpha\sqrt{p_{\rm gas}p_{\rm tot}}$). In all the
calculations,  $\alpha=0.5$ is adopted. The dotted lines represent
the spectra of the cold discs, while the dashed and dash-dotted
lines are for the Compton and synchrotron$+$bremsstrahlung radiation
respectively. The accretion rates: $\dot{m}=0.5$(red), $0.05$(green)
and $0.005$(blue) are adopted respectively(from up to down). The two
black dotted lines represent 2 and 10~keV, respectively. }
\label{fig3}
\end{figure}

\begin{figure}
\centerline{\psfig{figure=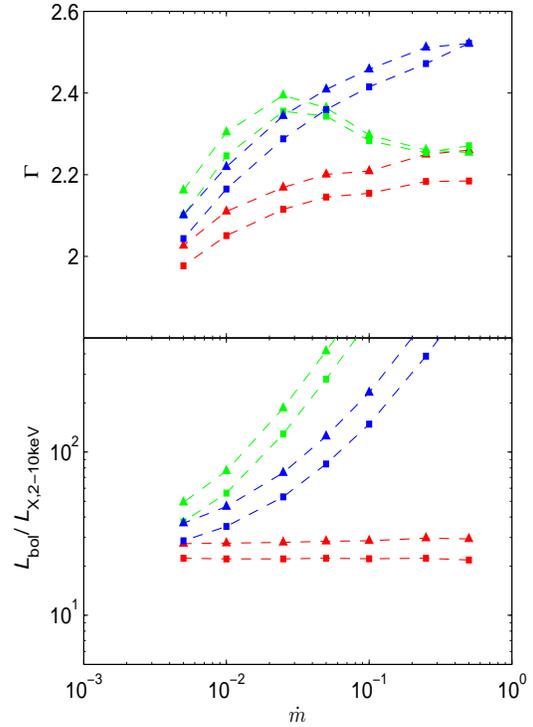,width=7.0cm,height=9.8 cm}}
\caption{The upper panel: the photon spectral indices as functions
of accretion rate $\dot{m}$ for different magnetic stress models
(red: $\tau_{r\varphi}=\alpha{p}_{\rm tot}$; green:
$\tau_{r\varphi}=\alpha{p}_{\rm gas}$; blue:
$\tau_{r\varphi}=\alpha\sqrt{p_{\rm gas}p_{\rm tot}}$). The squares
represent the results calculated with $\alpha=0.5$, while the
triangles are for $\alpha=0.3$. The X-ray bolometric correction
factors $L_{\rm bol}/L_{\rm X,2-10keV}$ as functions of accretion
rate $\dot{m}$ for different models are plotted in the lower panel.
} \label{fig4}
\end{figure}

In Fig. \ref{fig3}, we plot the spectra of the disc-corona systems
calculated with different magnetic stress tensors. The hard X-ray
emission indeed exhibits a power-law feature for all the models. The
photon spectral indices and the ratio of the bolometric luminosity
to the X-ray luminosity in 2--10~keV as functions of accretion rate
for different magnetic stress models are plotted in Fig. \ref{fig4}.
The photon indices $\Gamma$ do not change much with accretion rate
$\dot{m}$ for the disc-corona models with either
$\tau_{r\varphi}=\alpha{p}_{\rm tot}$ or
$\tau_{r\varphi}=\alpha{p}_{\rm gas}$, while the hard X-ray spectral
index $\Gamma$ increases significantly with accretion rate $\dot{m}$
for the model with $\tau_{r\varphi}=\alpha \sqrt{p_{\rm gas}p_{\rm
tot}}$.

\begin{figure}
\centerline{\psfig{figure=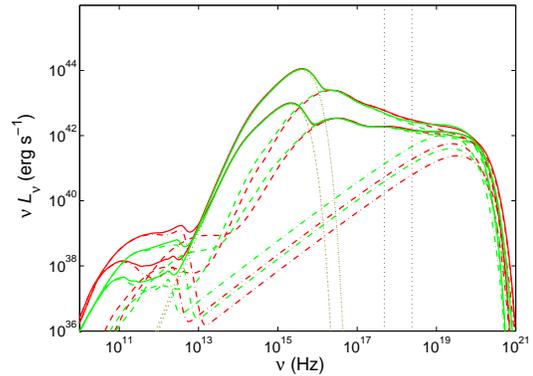,width=7.0cm,height=5.0 cm}}
\caption{The spectra of the disc corona systems calculated for
$\dot{m}=0.01$, and $0.1$, respectively
($\tau_{r\varphi}=\alpha\sqrt{p_{\rm gas}p_{\rm tot}}$ is adopted in
the calculations). The red lines represent the spectra for $T_{\rm
i}=0.9T_{\rm vir}$, while the green lines are for $T_{\rm
i}=0.3T_{\rm vir}$. The dotted lines represent the spectra of the
cold discs, while the dashed and dash-dotted lines are for the
Compton and synchrotron$+$bremsstrahlung radiation respectively.  }
\label{fig5}
\end{figure}

\begin{figure}
\centerline{\psfig{figure=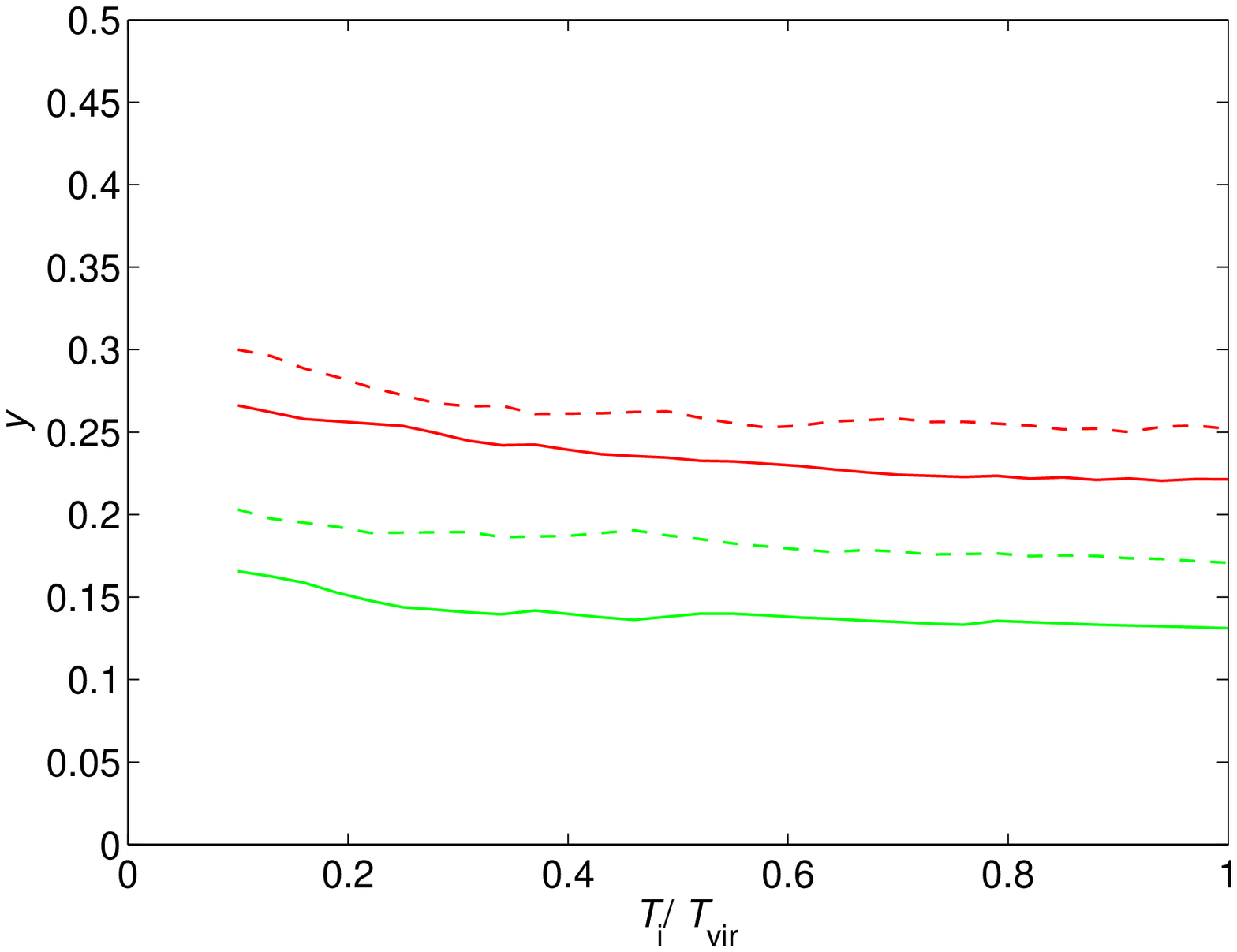,width=7.0cm,height=7.0 cm}}
\caption{The Compton $y$-parameter versus the ion temperature
$T_{\rm i}$ in the corona at radius $R$ with different accretion
rates $\dot{m}=0.01$(red), and $0.1$(green). The solid lines are
calculated with $R=10R_{\rm S}$, while the dashed lines are for
$R=40R_{\rm S}$. } \label{fig6}
\end{figure}

{We compare the spectra of the accretion disc/corona systems with
different ion temperature $T_{\rm i}$ adopted in Fig. \ref{fig5}. It
is found that the X-ray spectra change little with the value of
$T_{\rm i}$, provided all other parameters are fixed. In Fig.
\ref{fig6}, we plot the Compton $y$-parameter ($y=4kT_{\rm
e}\tau_{0}/m_{\rm e}c^{2}$) varying with $T_{\rm i}$, and find that
it is insensitive to $T_{\rm i}$.}

\section{Discussion}

It is believed that the power generated in the disc is transported
vertically with the buoyancy of the magnetic fields in the disc, and
the fraction of the power dissipated in the corona to the total is
mainly regulated by the magnetic fields
\citep*[e.g.,][]{d98,mf02,wwm04}. \citet{vf07} found that the
fraction $L_{\rm cor}/L_{\rm bol}$ is about 0.5 for the sources with
low Eddington ratio $L_{\rm bol}/L_{\rm Edd}\sim 0.01$, while it
decreases to 0.1 for $L_{\rm bol}/L_{\rm Edd}\sim 1$, for a sample
of AGN. From Fig. \ref{fig1}, we find that the model with
$\tau_{r\varphi}=\alpha p_{\rm tot}$ always predict a constant
$L_{\rm cor}/L_{\rm bol}$, which seems to be inconsistent with the
observation. The models with other two magnetic stress tensors can
roughly reproduce the trend of $L_{\rm cor}/L_{\rm Edd}$ decreasing
with accretion rate $\dot{m}$, though the model with
$\tau_{r\varphi}=\alpha p_{\rm gas}$ always underpredicts the
$L_{\rm cor}/L_{\rm bol}$ at high Eddington ratio end, even if
$\alpha=1$ is adopted. It seems that the model with
$\tau_{r\varphi}=\alpha \sqrt{p_{\rm gas}p_{\rm tot}}$ is better
than the other two models.

The ratio $L_{\rm cor}/L_{\rm bol}$ decreases with accretion rate
$\dot{m}$, which means the fraction of the power radiated from the
cold disc increases with $\dot{m}$. The cooling mechanism of the hot
corona is dominated by the inverse Compton scattering of the soft
photons from the cold disc by the hot electrons in the corona. There
are more soft photons supplied by the cold disc for high-$\dot{m}$
cases, and the fraction of the power radiated from the corona
decreases with $\dot{m}$. Thus, it can be easily understood that the
temperature of the hot electrons in the corona should decrease with
accretion rate $\dot{m}$ (see Fig. \ref{fig2}). It is found that the
temperature of the electrons in the corona decreases to $\sim
3-4\times 10^8$~K for the models with
$\tau_{r\varphi}=\alpha{p}_{\rm gas}$ or $\tau_{r\varphi}=\alpha
\sqrt{p_{\rm gas}p_{\rm tot}}$ when $\dot{m}=0.5$, which implies
that the hot corona is nearly to be suppressed when $\dot{m}$ is as
high as $\ga 0.5$.

We calculate the spectra of the accretion disc-corona systems, and
find that their hard X-ray continuum emission indeed exhibits a
power-law spectrum (see Fig. \ref{fig3}). In Fig. 4, we plot the
photon spectral indices $\Gamma$ in 2--10~keV and the X-ray
correction factors $L_{\rm bol}/L_{\rm X,2-10keV}$ as functions of
accretion rate $\dot{m}$ with different magnetic stress tensors. We
find that the photon spectral index $\Gamma$ predicted by the model
with $\tau_{r\varphi}=\alpha{p}_{\rm tot}$ is within $\sim 2-2.2$
for different values of $\dot{m}$, and it increases slightly with
$\dot{m}$. The X-ray correction factor $L_{\rm bol}/L_{\rm
X,2-10keV}$ remains constant with accretion rate $\dot{m}$ for this
model. The photon spectral index $\Gamma$ increases with $\dot{m}$
while $\dot{m}\la 0.02$, and it then decreases with $\dot{m}$, for
the model with $\tau_{r\varphi}=\alpha{p}_{\rm gas}$. We find that
the bremsstrahlung emission {contributes some to} the hard X-ray
energy band when $\dot{m}$ is high, and the photon spectral index
$\Gamma$ is {slightly affected by the bremsstrahlung emission} (see
the middle panel of Fig. 3). We find that the X-ray correction
factor $L_{\rm bol}/L_{\rm X,2-10keV}$ for this model becomes
extremely high for high-$\dot{m}$ cases, which seems to be
inconsistent with that derived from the observations \citep*[see
Fig. 12 in][]{vf07}. The X-ray spectra of the disc-corona systems
with $\tau_{r\varphi}=\alpha \sqrt{p_{\rm gas}p_{\rm tot}}$ show
that both the photon spectral index $\Gamma$ and the X-ray
correction factor $L_{\rm bol}/L_{\rm X,2-10keV}$ increase with
accretion rate $\dot{m}$, which are qualitatively consistent with
the observations, except for the low $\dot{m}$ end. The temperature
of the electrons in the corona decreases with $\dot{m}$ (see Fig.
\ref{fig2}), and the inverse Compton scattered X-ray spectrum
therefore becomes softer for a higher $\dot{m}$. {Similarly,
anticorrelations between the X-ray spectral indices and X-ray fluxes
were found in some X-ray binaries \citep*[e.g.,][]{y03,z04,d08},
which seem to be roughly consistent with our present disc corona
model. }

The photon spectral index $\Gamma$ is found to increase with the
Eddington ratio, and it could be as low as $\sim 1$ when
$\dot{m}\sim 0.01$ \citep*[see][]{s06,s08}. The hard X-ray photon
spectral index $\Gamma\sim 2$ derived from our model calculations
with any magnetic stress tensor when $\dot{m}\sim 0.01$, which are
significantly higher than the observed $\Gamma\sim 1$
\citep*[e.g.][]{s06,s08}. {Although the patchy corona model proposed
by \citet{hmg94} can produce the observed hard X-ray spectra with
$\Gamma\sim 1$, this patchy corona model is unable to explain the
correlation between the Compton reflection and the hard X-ray
spectral index. In an alternative model, the hot plasma above the
cold disc is assumed to move away from the cold disc at a mild
relativistic velocity \citep{b99}, which reduces both the reflection
and reprocessing in the cold disc. This naturally leads to harder
X-ray spectrum  (see the discussion in Sect. 1). However, the
detailed physics for producing such mildly relativistic outflows is
still unclear.}

It was argued that the central engines in these low-luminosity
sources with very hard X-ray emission may be different from their
high-luminosity counterparts, i.e., the advection dominated
accretion flows (ADAFs) may be present in these low-luminosity
sources \citep{ly99}. The soft incident photons for Comptonization
are mainly due to the synchrotron$+$bremsstrahlung emission in the
ADAFs. The energy density of the soft photons in the ADAF is much
lower than that in the disc-corona model, which leads to inefficient
cooling and relatively higher electron temperature in the ADAFs
\citep*[e.g.,][]{ny95}. Thus, the X-ray spectra of the ADAFs can be
much harder than those of the disc-corona systems. The very hard
X-ray spectra observed in these low luminosity sources can be well
modelled with the ADAFs accreting at low rates \citep*[see
e.g.,][for the spectral modeling for the hard X-ray emission from
the low luminosity AGN with ADAFs]{q99,xc08}. Our present model
calculations are limited to the disc-corona model, in which the cold
discs extend to the marginal stable orbits. The model of a
disc-corona connecting with inner ADAF may resolve the photon index
discrepancy at the low $\dot{m}$ end. {The geometry of this
ADAF$+$disc/corona scenario is quite similar to the hot
plasma$+$cold disc model proposed by \citet{zls99}. When the
accretion rate decreases to a critical value $\dot{m}_{\rm crit}$,
the inner cold disc may be truncated at radius $R_{\rm tr}$ and it
transits to an ADAF within this radius. The transition radius
$R_{\rm tr}$ may be close to the marginal stable orbits soon after
the accretion mode transition \citep*[e.g.,][]{q99,yn04,xc08}, and
then it may increase with decreasing accretion rate $\dot{m}$
\citep*[e.g.,][]{l99,rc00,sd02,yn04}. The X-ray spectrum of such an
ADAF$+$disc/corona system consists of emission from the inner ADAF
and outer corona. The ratio of the X-ray emission from the ADAF to
that from the corona increases with decreasing $\dot{m}$, and
therefore the photon spectral index $\Gamma$ may decrease smoothly
with decreasing  $\dot{m}$ provided the initial truncated radius of
the cold disc is not very large compared with  the marginal stable
orbits. Similar to \citet{zls99}'s model, the correlation between
the Compton reflection and the hard X-ray spectral index can also be
naturally explained by this ADAF$+$disc/corona model, if the
truncated radius $R_{\rm tr}$ increases with decreasing accretion
rate $\dot{m}$. This model can also successfully explain the
spectral behaviours of X-ray binaries \citep*[see][for a review and
references therein]{dgk07}. } The detailed calculations on such
ADAF$+$corona systems will be reported in our future work.

In this work, we simply assume $T_{\rm i}=0.9T_{\rm vir}$, motivated
by the previous work on the disc-corona model calculations
\citep{lms02,lmo03}. {We also check how $T_{\rm i}$ may affect our
results by tuning the value of $T_{\rm i}$, and find that the X-ray
spectra of the disc/corona systems change very little if all other
disc parameters are fixed (see Fig. \ref{fig5}). The cooling of the
corona is dominated by the inverse Compton radiation, which is
roughly proportional to Compton $y$-parameter. Thus, it is not
surprising that our calculations show the Compton $y$-parameter
varying little with $T_{\rm i}$ if all other parameters are fixed
(see equation \ref{cor_energy} and Fig. \ref{fig6}). The thickness
of the corona is mainly regulated by the ion temperature $T_{\rm
i}$, and therefore the electron density decreases with increasing
$T_{\rm i}$, which leads to significant difference in
synchrotron/bremsstrahlung spectra for different $T_{\rm i}$ (see
Fig. \ref{fig5}). } We find that the resulted $\dot{m}-\Gamma$
relations are almost not changed if a different value of $T_{\rm i}$
is adopted.
Unlike the ADAFs, almost all the power dissipated in the hot corona
with magnetic reconnection is radiated away locally. This means the
radiated power in the corona is independent of the value $\delta$,
and the temperature and density of the electrons in the corona are
almost insensitive with the value of $\delta$. We also perform the
same model calculations for different black hole masses (e.g.,
$M_{\rm bh}=10^9{\rm M}_\odot$). It is found that our results are
almost independent of $M_{\rm bh}$ for massive black holes.

In all our calculations, we assume the magnetic fields to be
equipartitioned with the gas pressure in the corona. As the cooling
of the corona is dominated by the inverse Comptonization of the soft
photons from the cold disc, the structure of the disc-corona and its
spectrum (except in the radio wavebands) is almost independent of
the magnetic field strength in the corona. Recently, \citet{lb08}
found there is a strong correlation between the radio luminosity
($L_{\rm R}$) and X-ray luminosity ($L_{\rm X}$) with $L_{\rm R}\sim
10^{-5}L_{\rm X}$, for the radio quiet Palomar-Green (PG) quasar
sample. The spectra of our disc-corona model show that the radio
emission is correlated with the X-ray emission, which is roughly
consistent with the correlation between $L_{\rm R}$ and $L_{\rm X}$
discovered by \citet{lb08}.

\section*{Acknowledgments}
I thank the referee, Zdziarski A.~A., for his helpful
suggestions/comments,  B.~F. Liu, T.~G. Wang, Q.~W. Wu and W. Yuan
for helpful discussion. This work is supported by the NSFC (grants
10773020, 10821302 and 10833002), the CAS (grant KJCX2-YW-T03), and
the National Basic Research Program of China (grant 2009CB824800).

\end{document}